\documentstyle[epsf,12pt]{article}

\renewcommand{\baselinestretch}{1.65}

\begin{document}

\title{GEODESICS IN THE $\gamma$ SPACETIME}
\author{L. Herrera$^1$\thanks{Postal address: Apartado 80793, Caracas 1080A,
Venezuela; e-mail: laherrera@telcel.net.ve}
,\ \
Filipe M. Paiva$^2$\thanks{e-mail: fmpaiva@symbcomp.uerj.br}
\\ \ \ and \ \
N. O. Santos$^3$\thanks{e-mail: nos@lacesm.ufsm.br}
\\ \\
{\small $^1$Escuela de F\'{\i}sica, Facultad de Ciencias,}\\
{\small Universidad Central de Venezuela, Caracas, Venezuela, and}\\
{\small Centro de Astrof\'{\i}sica Te\'orica, Merida, Venezuela.}
\\ \\
{\small $^2$Departamento de F\'{\i}sica Te\'orica, Universidade do Estado do
Rio de Janeiro,}\\
{\small Rua S\~ao Francisco Xavier 524, 20550-013 Rio de Janeiro - RJ,
Brazil.}
\\ \\
{\small $^3$Laborat\'orio de Astrof\'{\i}sica e Radioastronomia,}\\
{\small Centro Regional Sul de Pesquisas Espaciais - INPE/MCT}\\
{\small LACESM - Cidade Universit\'aria, 97105-900 Santa Maria RS, Brazil.}}
\date{}

\renewcommand{\baselinestretch}{1.1}
\maketitle

\begin{abstract}
Circular and radial geodesics are studied in the spacetime described by the
$\gamma$ metric. Their behaviour is compared with the spherically symmetric
situation, bringing out the sensitivity of the trajectories to deviations
from spherical symmetry.
\end{abstract}

\newpage

\section{Introduction} 

The influence of small perturbations of a Schwarzschild black hole, on the
trajectories of test particles around the source, has atracted the attention
of many researchers. These perturbations are usually introduced as
additional mass and charge concentrations
\cite{Chandrasekhar,Contopoulos,Moeckel,Dettmann,Vieira,Cornish}, magnetic
fields \cite{Karas} or gravitational waves \cite{Bombelli}. The common
results of all these works being, that essentially any perturbation of a
Schwarzschild black hole will lead to chaotic orbits. \par

In this work we want to present another approach to the problem of
introducing perturbations to spherical symmetry. This consists in
considering an exact solution of Einstein equations continuously linked to
the Schwarzschild metric, through one of its parameters. The rationale
behind this approach lies on the known \cite{Winicour}, though usually
overlooked, fact that as the source approaches the horizon, any finite
perturbation of the Schwarzschild spacetime, becomes fundamentally different
from any Weyl metric, even if the latter is characterized by parameters
whose values are arbitrarily close to those corresponding spherical
symmetry.\par

The solution to be considered here is the so called $\gamma$ metric
\cite{Esposito,Virbhadra}, which is also known as the Zipoy-Vorhees metric
\cite{Bach}, and belongs to the family of the Weyl solutions \cite{Weyl}.\par

The motivation for this choice may be found in the fact that the $\gamma$
metric corresponds to a solution of the Laplace equation, in cylindrical
coordinates, with the same Newtonian source image \cite{Bonnor}, as the
Schwarzschild metric (a rod). In this sense the $\gamma$ metric appears as
the {\it natural} generalization of the Schwarzschild spacetime to the
static axisymmetric case.\par

We shall find the geodesic equations for test particles in the $\gamma$
metric. Particular attention will be devoted to circular and radial
geodesics. The qualitative differences in the dynamics of the test particles
as compared to the spherically symmetric case will be illustrated and
discussed.\par

The paper is organized as follows. In the next section the $\gamma$ metric
is briefly presented. In section 3 geodesic equations are found and analyzed
and in section 4 the gravitational and centrifugal forces are studied.
Finally the results are discussed in the last section.

\section{The $\gamma$ metric}  

In cylindrical coordinates, static axisymmetric solutions to the Einstein
equations are given by the Weyl metric \cite{Weyl}
\begin{equation}
ds^2=e^{2\lambda}dt^2-e^{-2\lambda}[e^{2\mu}(d\rho^2+dz^2)+\rho^2d\phi^2],
\end{equation}
with
\begin{equation}
\lambda_{,\rho\rho}+\rho^{-1}\lambda_{,\rho}+\lambda_{,zz}=0,
\end{equation}
and
\begin{equation}
\mu_{,\rho}=\rho(\lambda^2_{,\rho}-\lambda^2_{,z}), \qquad
\mu_{,z}=2\rho\lambda_{,\rho}\lambda_{,z}.
\end{equation}
Observe that (2) is just the Laplace equation for $\lambda$ in the Euclidean
space.
\par
The $\gamma$ metric is defined by \cite{Esposito}
\begin{eqnarray}
e^{2\lambda}&=&\left(\frac{R_1+R_2-2m}{R_1+R_2+2m}\right)^\gamma,\\
e^{2\mu}&=&\left[\frac{(R_1+R_2+2m)(R_1+R_2-2m)}{4R_1R_2}\right]^{\gamma^2},
\end{eqnarray}
where
\begin{equation}
R_1^2=\rho^2+(z-m)^2, \qquad R_2^2=\rho^2+(z+m)^2.
\end{equation}
It is worth noticing that $\lambda$, as given by (4), corresponds to the
Newtonian potential of a line segment of mass density $\gamma/2$ and length
$2m$, symmetrically distributed along the $z$ axis. The particular case
$\gamma=1$, corresponds to the Schwarzschild metric.
\par
It will be useful to work in Erez-Rosen coordinates \cite{Bach}, given by
\begin{equation}
\rho^2=(r^2-2mr)\sin^2\theta, \qquad z=(r-m)\cos\theta,
\end{equation}
which yields the line element (1) with (4,5) as \cite{Esposito}
\begin{equation}
ds^2=Fdt^2-F^{-1}[Gdr^2+Hd\theta^2+(r^2-2mr)\sin^2\theta d\phi^2],
\end{equation}
where
\begin{eqnarray}
F&=&\left(1-\frac{2m}{r}\right)^\gamma,\\
G&=&\left(\frac{r^2-2mr}{r^2-2mr+m^2\sin^2\theta}\right)^{\gamma^2-1},\\
H&=&\frac{(r^2-2mr)^{\gamma^2}}{(r^2-2mr+m^2\sin^2\theta)^{\gamma^2-1}}.
\end{eqnarray}
Now it is easy to check that $\gamma=1$ corresponds to the Schwarzschild
metric.
\par
The total mass $M$ of the source is \cite{Esposito,Virbhadra}
\begin{equation}
M=\gamma m,
\end{equation}
and the quadrupole moment $Q$ is given by
\begin{equation}
Q=\frac{\gamma}{3}(1-\gamma^2)M^3.
\end{equation}
{}From (13) we have that $\gamma>1 (\gamma<1)$ corresponds to the oblate
(prolate) spheroid.

\section{The geodesics}

The equations governing the geodesics can be derived from the Lagrangian
\begin{equation}
2{\cal L}=g_{\alpha\beta}\dot{x}^\alpha\dot{x}^\beta,
\end{equation}
where the dot denotes differentiation with respect to an affine parameter
$s$ , which for timelike geodesics coincides with the proper time. Then,
from the Euler-Lagrange equations,
\begin{equation}
\frac{d}{ds}\left(\frac{\partial{\cal
L}}{\partial\dot{x}^\alpha}\right)-\frac{\partial{\cal L}}
{\partial x^\alpha}=0,
\end{equation}
it follows, for the metric (8),
\begin{eqnarray}
\ddot{t}+\gamma\dot{t}\dot{r}\nu^\prime=0,
\\ \mbox{}\nonumber\\
-2\ddot{r}\frac{e^{-\gamma\nu}}{A^{\gamma^2-1}}+\dot{r}^2
\left[\gamma\frac{e^{-\gamma\nu}\nu^\prime}{A^{\gamma^2-1}}+
(1-\gamma^2)m^2\:\frac{e^{-(1+\gamma)\nu}}
{r^2A^{\gamma^2}}\left(\frac{2}{r}+\nu^\prime\right)\sin^2\theta
\right] \nonumber \\
-4\dot{r}\dot{\theta}(1-\gamma^2)m^2\:\frac{e^{-(1+\gamma)\nu}}
{r^2A^{\gamma^2}}\sin\theta\cos\theta-\dot{t}^2\gamma\nu^\prime
e^{\gamma\nu} \nonumber \\
+\dot{\theta}^2\left\{
\frac{r^2e^{(1-\gamma)\nu}}{A^{\gamma^2-1}}\left[\frac{2}{r}+(1-\gamma)
\nu^\prime\right]+(1-\gamma^2)m^2\:\frac{e^{-\gamma\nu}}
{A^{\gamma^2}}\left(\frac{2}{r}+\nu^\prime\right)\sin^2\theta\right\}
\nonumber \\
+\dot{\phi}^2r^2e^{(1-\gamma)\nu}\left[\frac{2}{r}+
(1-\gamma)\nu^\prime\right]\sin^2\theta=0,
\\ \mbox{}\nonumber\\
-\ddot{\theta}\frac{r^2e^{(1-\gamma)\nu}}{A^{\gamma^2-1}}
-\dot{\theta}^2(1-\gamma^2)m^2\:\frac{e^{-\gamma\nu}}
{A^{\gamma^2}}\sin\theta\cos\theta\nonumber \\
+\dot{\theta}\dot{r}\left\{-\frac{r^2e^{(1-\gamma)\nu}}{A^{\gamma^2-1}}
\left[\frac{2}{r}+(1-\gamma)\nu^\prime\right]+(1-\gamma^2)m^2\:
\frac{e^{-\gamma\nu}}{A^{\gamma^2}}\left(\frac{2}{r}+\nu^\prime\right)
\sin^2\theta
\right\}\nonumber \\
+\dot{r}^2(1-\gamma^2)m^2\:\frac{e^{-(1+\gamma)\nu}}
{r^2A^{\gamma^2}}\sin\theta\cos\theta+\dot{\phi}^2r^2e^{(1-\gamma)\nu}
\sin\theta\cos\theta=0,
\\ \mbox{} \nonumber \\
\left\{\ddot{\phi}+\dot{\phi}\dot{r}\left[\frac{2}{r}+(1-\gamma)
\nu^\prime\right]+2\dot{\phi}\dot{\theta}\cot\theta\right\}
e^{(1-\gamma)\nu}\sin^2\theta=0,
\end{eqnarray}
with
\begin{equation}
e^\nu\equiv 1-\frac{2m}{r}, \qquad A\equiv
1+\frac{m^2e^{-\nu}\sin^2\theta}{r^2},
\end{equation}
and prime denotes differentiation with respect to $r$.It is a simple matter
to check that if $\gamma=1$ we recover the geodesic equations of the
Schwarzschild spacetime.
\par
Let us first consider circular geodesics. From (16-19) we obtain using
$\dot{r}=\dot{\theta}=0$,
\begin{eqnarray}
\ddot{t}=\ddot{\phi}&=&0,\\
\gamma\dot{t}^2e^{\gamma\nu}\nu^\prime&=&
\dot{\phi}^2r^2e^{(1-\gamma)\nu}
\left[\frac{2}{r}+(1-\gamma)\nu^\prime\right]\sin^2\theta,\\
2\dot{\phi}^2r^2e^{(1-\gamma)\nu}\sin\theta\cos\theta&=&0.
\end{eqnarray}
Then from the definiton of the angular velocity $\omega$ of a test particle
along circular geodesics
\begin{equation}
\omega=\frac{\dot{\phi}}{\dot{t}},
\end{equation}
we obtain, using (22),
\begin{equation}
\omega^2=\frac{\gamma m}{[r-(1+\gamma)m]r^2\sin^2\theta}\left(1-\frac{2m}{r}
\right)^{2\gamma-1}.
\end{equation}
First of all, we observe that, as it follows from (23), circular geodesics
out of the equatorial plane $\theta=\pi/2$, are now possible, if only
$\gamma<1$ and $r=2m$. This kind of trajectories do not exist in
Schwarzschild spacetime ($\gamma=1$). {}From (25) it also follows that
physically meaningful values of $\omega$ at $r=2m$, exist only for
$\gamma=1/2$. Therefore circular geodesics for $\theta\neq\pi/2$ can ocur if
only $\gamma=1/2$ and $r=2m$, with angular velocity
\begin{equation}
\omega^2=\frac{1}{4m^2\sin^2\theta}.
\end{equation}
Circular geodesics outside the equatorial plane also exist in the Kerr
metric
\cite{Bonnor1}. These kind of orbits implies the presence of repulsive
forces whose nature is still not well understood \cite{Schucking, Herrera}.
However it should be stressed that since $r=2m$ represents a physical
singularity in the $\gamma$ metric, these orbits are deprived of physical
meaning. In the weak field limit one obtains from (25), for $\theta=\pi/2$,
up to the first order in $m/r$,
\begin {equation}
\omega^2\approx\frac{\gamma m}{r^3}\left[1+(1-\gamma)\frac{3m}{r}\right]+
O\left(\frac{m}{r}\right)^2.
\end{equation}
If $\gamma=1$, we recover the well known Kepler law, which is also valid, if
$\gamma=1$, without any approximation.
\par
\begin{figure}[p]
\epsffile{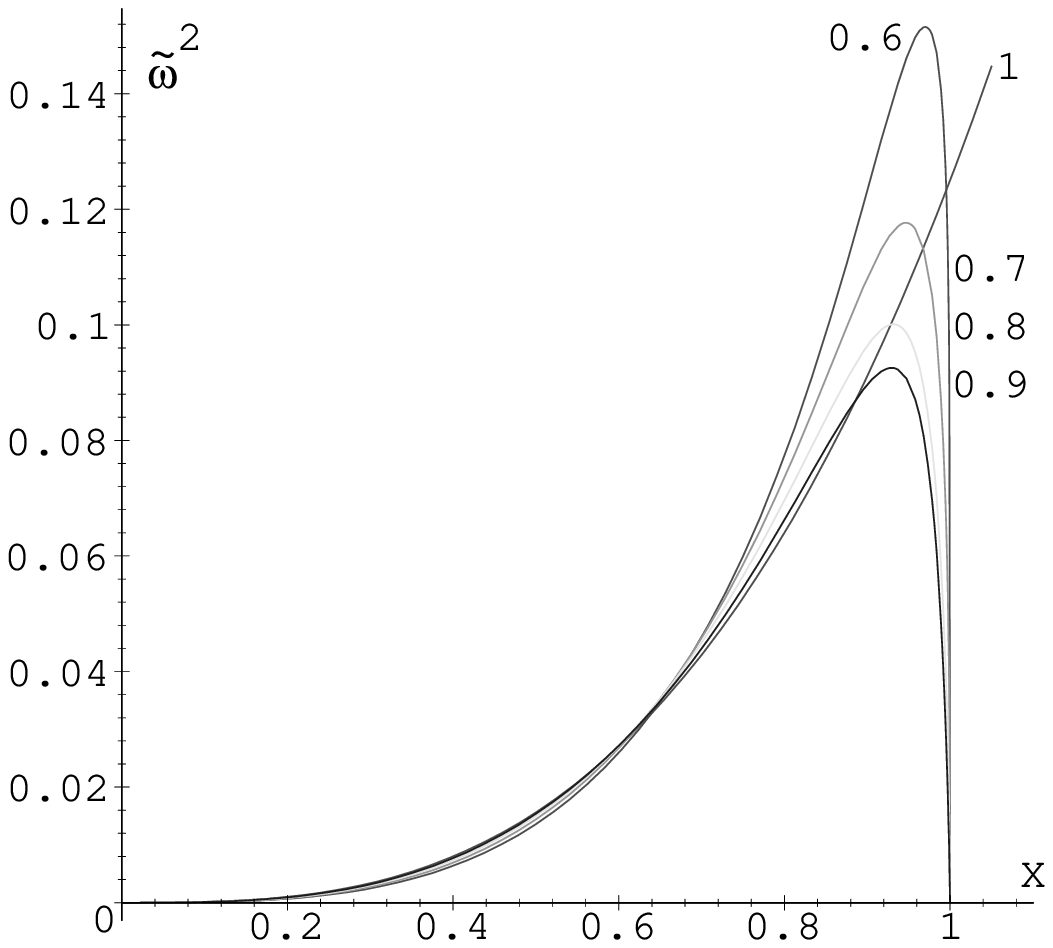}
\caption{$\tilde\omega^2$ as function of $x$ for the five values of $\gamma$ 
indicated.}
\label{figure1}
\end{figure}

If $\gamma\neq 1$ the second term within the brackets, in (27), gives the
correction due to the quadrupole moments of the source. In troducing the
dimensionless variable $x\equiv 2m/r$, we can rewrite (25), for
$\theta=\pi/2$,
\begin {equation}
\tilde{\omega}^2\equiv m^2\omega^2=\frac{\gamma(1-x)^{2\gamma-1}x^3}
{4[(1-\gamma)x+2(1-x)]}.
\end{equation}
Figure~\ref{figure1} 
shows the behaviour of $\tilde{\omega}^2$ as a function of $x$ for
different values of $\gamma$, including $\gamma=1$.
\par
The bifurcation between the $\gamma=1$ and $\gamma\neq 1$ cases, for values
of $x$ close to one, clearly illustrates the sensitivity of the system under
perturbations of $\gamma$ in the neighborhood of $\gamma=1$.
\par
\begin{figure}[p]
\epsffile{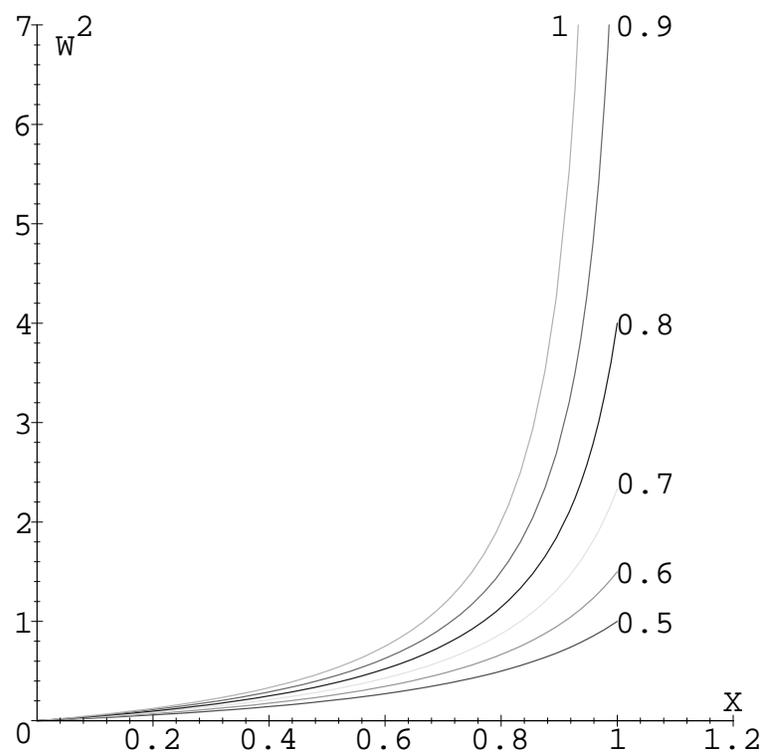} 
\caption{$W^2$ as function of $x$ for six different values of $\gamma$.}
\label{figure2}
\end{figure}

For comparative purposes it will be useful to find an expression for the
tangential velocity $W^\mu$ of the test particle along the circular
geodesic.
{}From \cite{Anderson}, we have
\begin{equation}
W^\alpha=\frac{V^\alpha}{\sqrt{g_{00}}dx^0},
\end{equation}
with
\begin{equation}
V^\alpha=(0,dx^1,dx^2,dx^3),
\end{equation}
and we obtain from (8) for $\theta=\pi/2$,
\begin{equation}
W^2\equiv W_\alpha
W^\alpha=\left(1-\frac{2m}{r}\right)^{1-2\gamma}\omega^2r^2=\frac{\gamma m}
{r-(1+\gamma)m},
\end{equation}
where $\omega^2$ is given by (25). In the weak field limit, $m/r\ll 1$, the
{\it classical} expression, $W=\omega r$ is recovered. Figure~\ref{figure2} 
shows $W^2$
as a function of $x$ for different values of $\gamma$. Furthermore, we see
from (31), that null circular geodesics appear when $r=(1+2\gamma)m$. Since
the physical singularity lies for $r=2m$ then null circular geodesics can
only exist for $\gamma>1/2$, otherwise, for $\gamma<1/2$, they do not exist.
\par
The Levi-Civita metric, which represents the field of an infinite line mass
of constant energy density $\sigma$, can be obtained as a limiting case from
the $\gamma$ metric by taking $m\rightarrow\infty$ and associating
$\gamma=2\sigma$ \cite{Herrera1}. Next we see, in the equatorial plane, how
the angular velocity $\omega$ and the tangential velocity $W$ of the
circular geodesics, respectively given by (25) and (31), behave for big
values of $m$.
\par
Rewriting (25) in terms of the coordinates given by (7), with
$\theta=\pi/2$, we obtain
\begin{equation}
\omega^2=\frac{\gamma
m(\sqrt{m^2+\rho^2}-m)^{2\gamma-1}}{(\sqrt{m^2+\rho^2}-\gamma m)
(m+\sqrt{m^2+\rho^2})^{2\gamma+1}}.
\end{equation}
Expanding in series (32) for big values of $m$ and keeping only the term in
its lowest order of $\rho/m$, we obtain
\begin{equation}
\omega^2\approx\frac{2\sigma}{1-2\sigma}(4m^2)^{-4\sigma}\rho^{2(4\sigma-1)}
,
\end{equation}
where we have substituted $\gamma=2\sigma$. The expression (33) is the same
as for the Levi-Civita metric obtained in \cite{Herrera} provided the
topological defect $a$, as given in \cite{Herrera}, is associated in (33)
as $a=(4m)^{-2\sigma}$. In \cite{Herrera1} it is proved that the limiting
case of Levi-Civita spacetime from the $\gamma$ spacetime produces an
infinite topological defect.
\par
Now rewriting (31) in terms of the coordinates given by (7), we have
\begin{equation}
W^2=\frac{\gamma m}{\sqrt{m^2+\rho^2}-\gamma m}.
\end{equation}
Expanding (34) in series for big values of $m$ up to the second order of
$\rho/m$, we obtain
\begin{equation}
W^2\approx\frac{2\sigma}{1-2\sigma}-\frac{2\sigma}{2(1-2\sigma)^2}
\left( \frac{\rho}{m}\right)^2+O\left(\frac{\rho}{m}\right)^4,
\end{equation}
where we have substituted $\gamma=2\sigma$. From (35) we have that, for a
given $\gamma$ and a fixed radius $\rho$, decreasing the length $2m$ of the
line segment of mass decreases the tangential speed $W$ of the circular
geodesics.When $m\rightarrow\infty$ we have that the tangential velocity
(34) is the same as the one obtained for the Levi-Civita metric
\cite{Herrera}, being
\begin{equation}
\lim_{m\rightarrow\infty}W^2=\frac{\gamma}{1-\gamma}=\frac{2\sigma}
{1-2\sigma}.
\end{equation}
Observe that (36) sets a constraint on possible values of $\sigma$ to avoid
values of $W$ larger than 1, i.e. the velocity of light. This is in contrast
with the situation in the $\gamma$ metric (34), where such constraint
$W\leq1$ involves $\gamma$ and the radial coordinate $\rho$ of the orbit.
\par
\begin{figure}[p]
\epsffile{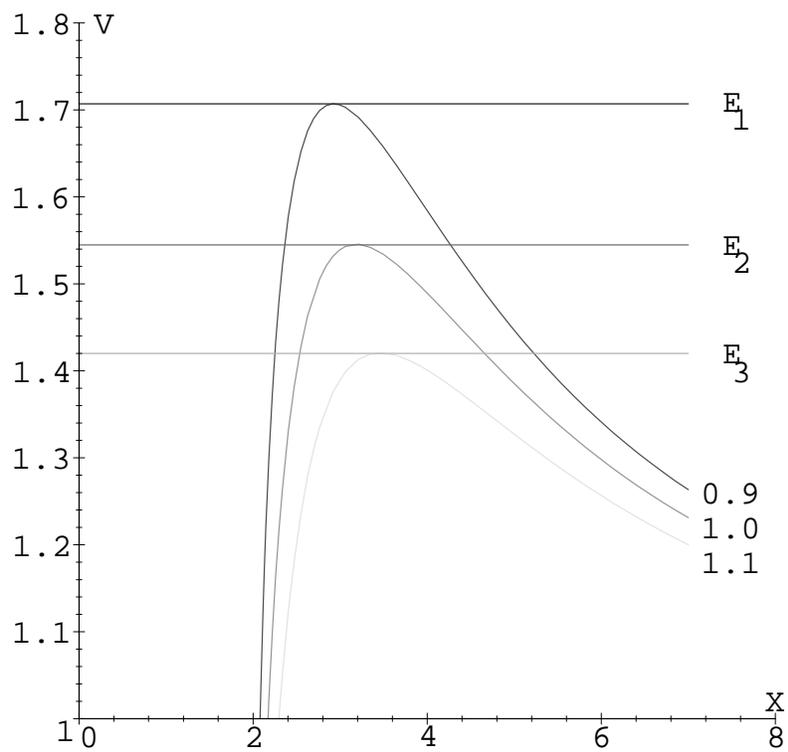}
\caption{$V$ as function of $x$ for three different values of $\gamma$. For
the three values of $E$, there are three different unstable circular
orbits.}
\label{figure3}
\end{figure}

Let us now consider radial geodesics. From (8), in the $\theta=\pi/2$ plane,
it follows
\begin{eqnarray}
1=\dot{t}^2\left(1-\frac{2m}{r}\right)^\gamma-\dot{r}^2\left(1-\frac{2m}{r}
\right)^{-\gamma}\left(\frac{r^2-2rmr}{r^2-2mr+m^2}\right)^{\gamma^2-1}
\nonumber \\
- \dot{\phi}^2\left(1-\frac{2m}{r}\right)^{1-\gamma}r^2.
\end{eqnarray}
Next, it follows from (15),
\begin{eqnarray}
\frac{\partial{\cal L}}{\partial\dot{t}}=\mbox{constant}=E &=& 
\dot{t}\left(1-\frac{2m}{r}\right)^\gamma,\\
\frac{\partial{\cal L}}{\partial\dot{\phi}}=\mbox{constant}=L &=& 
-\dot{\phi}\left(1-\frac{2m}{r}\right)^{1-\gamma}r^2,
\end{eqnarray}
where $E$ and $L$ represent, respectively, the total energy and the angular
momentum of the test particle. Then using (38,39) in (37), we obtain
\begin{equation}
\dot{r}^2=\left(1-\frac{2m}{r}\right)^{\gamma^2-1}(E^2-V^2),
\end{equation}
where $V(x)$, which can be associated to the potential energy of the test
particle, is given by
\begin{equation}
V^2\equiv\left(1-\frac{2m}{r}\right)^\gamma\left[\frac{m^2\tilde{L}^2}{r^2}
\left(1-\frac{2m}{r}\right)^{\gamma-1}+1\right],
\end{equation}
with
\begin{equation}
\tilde{L}^2\equiv \frac{L^2}{m^2},
\end{equation}
or, in terms of $x$ (41) becomes
\begin{equation}
V^2=\left(1-\frac{2}{x}\right)^\gamma\left[\frac{\tilde{L}^2}{x^2}
\left(1-\frac{2}{x}\right)^{\gamma-1}+1\right].
\end{equation}
Figure~\ref{figure3}
shows $V$ for different values of $\gamma$ and $\tilde {L}^2=55$.
For indicated values of the total energy $E$, there exist unstable circular
orbits, which illustrate the sensitivity to small perturbations.

\section{Gravitational and centrifugal forces}

The study of the gravitational and centrifugal forces follows the same
treatment as in \cite{Sonego}. The four velocity of a particle moving on a
circular orbit in the equatorial plane $\theta=\pi/2$ of the $\gamma$
spacetime (8) can be expressed as
\begin{equation}
v^\alpha=(\delta^\alpha_t+\Omega\delta^\alpha_\phi)\Gamma,
\end{equation}
where $\Omega$ is a parameter and, since $v_\alpha v^\alpha=1$, $\Gamma$ is
given by
\begin{equation}
\Gamma^2=\frac{F}{F^2-\Omega^2r(r-2m)}.
\end{equation}
Assuming uniform motion, $\Omega=$constant, the acceleration $a^\alpha$ of
the particle,
\begin{equation}
a^\alpha=v^\beta v^\alpha_{;\beta},
\end{equation}
using (8,44,45) is
\begin{eqnarray}
a^\alpha=\delta^\alpha_r\frac{r^{1-\gamma-\gamma^2}(r-2m)^{\gamma(1-\gamma)}
(r-m)^{2(\gamma^2-1)}}{r^{-2\gamma}(r-2m)^{2\gamma-1}-\Omega^2r}\nonumber \\
\times\left\{\frac{\gamma
m}{r^2}\left(1-\frac{2m}{r}\right)^{2\gamma-1}-\Omega^2[r-(1+\gamma)m]
\right\}.
\end{eqnarray}
If $\gamma=1$, then (47) reduces to the Schwarzschild spacetime result
\cite{Sonego},
\begin{equation}
a^\alpha=\delta^\alpha_r\frac{1}{r}\left(1-\frac{2m}{r}\right)\frac{m-
\Omega^2r^3}{r-2m-\Omega^2r^3}.
\end{equation}
The acceleration $a^\alpha$ can be split into its gravitational component
$g^\alpha$, when $\Omega=0$, and its centrifugal component $c^\alpha$, as
\begin{equation}
-a^\alpha=g^\alpha+c^\alpha,
\end{equation}
and so we have from (47),
\begin{eqnarray}
g^\alpha &=& -\delta^\alpha_r\frac{\gamma m}{r^2}
\left(1-\frac{2m}{r}\right)^{\gamma(1-\gamma)}\left(1-\frac{m}{r}
\right)^{2(\gamma^2-1)},\\
c^\alpha &=& \delta^\alpha_r\Omega^2r[r-(1+2\gamma)m]\frac{(r-2m)
^{\gamma(1-\gamma)}(r-m)^{2(\gamma^2-1)}}{r^{\gamma(\gamma-1)}(r-2m)
^{2\gamma-1}-\Omega^2r^{\gamma^2+\gamma+1}}.
\end{eqnarray}
When the gravitational acceleration is balanced by the centrifugal
acceleration, which means $a^\alpha=0$, then we have $\Omega=\omega$, as it
can be checked from (25). The gravitational acceleration $a^\alpha$ always
points inwards, being $g^\alpha<0$. The centrifugal acceleration $c^\alpha$
for values of $r>(1+2\gamma)m$ is always positive, while for
$r<(1+2\gamma)m$ it becomes negative. Since $r=2m$ is the physical
singularity only $1/2<\gamma$ allows negative values for $c^\alpha$, while
$1/2>\gamma$ makes always $c^\alpha$ positive.

\section{Conclusions}

We have considered deviations from spherical symmetry by considering exact
solutions of the Weyl family, instead of perturbing the Schwarzschild
metric. As it should be expected from the Israel theorem \cite{Israel},
these approaches differ qualitatively as the orbit of the test particle gets
close to the horizon. Though it was not our purpose here to describe the
chaotic behaviour of the trajectories, it should be clear that such
behaviour is the expression of the sensitivity to small changes of $\gamma$.
This sensitivity in turn, appears sistematically in the cinematics of the
particles, for orbits close to $2m$, as illustrated by
figures~\ref{figure1}--\ref{figure3}.

\section*{Acknowledgment}

FMP gratefully acknowledges financial assistance from FAPERJ and the
Laborat\'orio de Astrof\'{\i}sica e Radioastronomia - Centro Regional Sul de
Pesquisas Espaciais (INPE/MCT - Santa Maria RS) for the kind hospitality
during the earlier stage of this work.

\end{document}